\documentclass[12pt]{article}

\usepackage[T1,OT1,TS1,T2A]{fontenc}
\usepackage{epsfig,amsmath,amsfonts,amssymb}

\usepackage{ifthen}

\def \beq {\begin{equation}}
\def \eeq {\end{equation}}
\def \beqar {\begin{eqnarray}}
\def \eeqar {\end{eqnarray}}
\def \beqa* {\begin{eqnarray*}}
\def \eeqa* {\end{eqnarray*}}

\def \d {\,{\rm{d}}}

\def\refart#1#2#3#4#5#6#7{{#1}, \textit{#3} \textbf{#4} {(#6)} {#5}{\ifthenelse{\equal{#7}{}}{}{ [#7]}}}%

\def\refbook#1#2#3#4#5#6#7{{#1}, \textit{#2}, #3, {\ifthenelse{\equal{#4}{}}{}{ch.~#4}}, #5, #6, #7}%

\begin{document}
\title{Isotropization in  Chaplygin matter universes\\ connected by a wormhole}%
\author{Anna~Mokeeva$^{1)}{}^\dag$ and Vladimir~Popov$^{2)}{}^\ddag$}%
\date{}

\maketitle%

{\small
\noindent$^{1)}$\textit{Physical Department, Lobachevsky State University of Nizhni Novgorod, Gagarin ave. 23, Nizhni Novgorod 603950, Russia}\\
$^{2)}$\textit{Institute of Physics, Kazan Federal University, Kremlyovskaya st. 18, Kazan 420008, Russia}\\
\\
$^\dag$anne.zaharovoy@gmail.com\\
$^\ddag$vladipopov@mail.ru
}

\begin{abstract}
Chaplygin anisotropic matter governing the cosmological evolution in
two identical universes with  an intermediate static spherically
symmetric region is considered. The static region contains a
wormhole allowing one to pass between two horizons. The metric in
the non-static region represents a kind of an anisotropic
Kantowski-Sachs cosmological model starting from a horizon instead
the initial singularity. A classification of the cosmologies with a
monotonic late stage is presented. It is shown that only one
scenario can involve a de Sitter regime. The scales in the de Sitter
phase allows one to describe the earliest accelerated expansion on
the classical level.
\end{abstract}

\vspace*{12pt}

{\small

}

\newpage

\section{Introduction}

One of the fundamental problems of the relativistic cosmology is the
problem of the initial cosmological singularity. The direct way to
solve the problem implies quantization of gravity. This line of
research develops due to efforts in the string theory, M-theory and
loop quantum gravity, and achievements in high-energy physics also.
An alternative view on the origin of the Universe supposes to
exclude the singularity in its geometrical sense. This approach is
undoubtedly stimulated by the extensive studying of the modern
accelerated expansion of the Universe\footnote{Note, that many of
cosmologies without the singularity was invented long before the
discovery of the present-day Universe acceleration (see
\cite{Starobinsky} for example) and has been incorporated into dark
energy models in an invariable or  modified form.} and involves dark
energy models such as the cosmological constant,  phantom fields,
exotic equations of state, etc., as well as a modified gravity. The
universe evolution without the singularity can be realized in some
ways. Among them are eternally existing universes including emergent
universes \cite{Ellis}, cyclic evolution \cite{kanekar} and bouncing
cosmologies \cite{Novello}. It is worth to mark out the ekpyrotic
scenario \cite{Khoury} because of its intriguing confrontation to
the inflationary paradigm.

One further feasibility to avoid the initial singularity is
universes expanding from a Killing horizon
\cite{Frolov,Dymnikova,Bronnikov1,Bronnikov2,Bronnikov3,Bronnikov4,Mokeeva}.
This approach arises from the idea that the universe can emerge in
an interior of a black hole. More than a half of a century  Gliner
\cite{Gliner} speculated that a de Sitter world can be result of a
gravitational collapse. Pathria \cite{Pathria}  proposed a more
definite model of a cosmic expansion beyond an event horizon. The
original attention to universes inside the black holes was
concentrated on the inquiry of geometrical properties and
cosmological details rather than on exact solutions of the Einstein
equations \cite{Frolov}. Models with a vacuum source providing a
regular black hole core were suggested in \cite{Dymnikova}.
Bronnikov et al. \cite{Bronnikov1} used the exact solutions for
phantom scalar fields to study and classify regular black holes
where singularity is replaced by a cosmological expansion and which
therefore were called \emph{black universes}. Such objects can be
supported by phantom matters and could arise from collapse in
another ambient universe. In the work \cite{Bronnikov2} the
Lema\^{\i}tre type cosmologies starting from a horizon for an
anisotropic perfect fluid with the vacuum equation of state were
considered for different topologies of the spacetime. The authors
suggested to call this type of the universe origin a \emph{null big
bang} (NBB) since a horizon is a null surface where the spatial
volume vanishes. A regularity conditions were studied for the NBB
cosmologies with neutral and charged matter obeying the barotropic
equation of state \cite{Bronnikov4}, and  matter with non-vacuum
behavior when the equation of state parameter $w=p_r/\rho\ne -1$
everywhere \cite{Bronnikov3}.

We construct a similar structure with a Chaplygin wormhole. This
kind of wormholes  can not give an asymptotic flatness and
encourages singularities \cite{Lobo1} in a completely static
spacetime. The wormhole  surrounded by two horizons naturally
avoided these disadvantages. The static region connects two
cosmological regions and the horizons appear as a NBB.  In
\cite{Mokeeva} we derived the regularity conditions for the metric
and the Chaplygin matter on the horizon and obtain some exact
solutions corresponding to different cosmological scenarios.

It was also shown that the considered spacetime configuration is
anisotropic in both, static and non-static, regions. The
observational data  are evidence that our Universe is isotropic to a
great extent. This does not forbid, however, an anisotropic behavior
at early stage. Our main motivation for this paper is to impart more
validity to the considered model  through conditions providing
isotropization after an anisotropic phase. We keep it in mind when
analysing and classifying cosmologies concerned.

The paper is organized as follows. In the next section we consider
the general properties of spacetimes with a spherically symmetric
wormhole surrounded by two horizons and external universes which are
supported by matter with the Chaplygin equation of state. The static
region is described in Sec.~\ref{Sec-Static region}. The
classification of cosmological scenarios is presented in
Sec.~\ref{Sec-ExactSolutions}. In  Sec.~\ref{Sec-Conclusion} we
summarize the results and make concluding remarks.

\section{Structure of the spacetime}
\label{Sec-Conditions}

We start our consideration with a static spherically symmetric spacetime containing an anisotropic perfect fluid in which an energy density and a radial pressure are related by Chaplygin's equation of state
\begin{equation} \label{Chap}
 p_{r}=-\dfrac{\alpha}{\rho},
\end{equation}
where $\alpha$ is a positive constant. The metric of the spacetime is taken in the form
\begin{equation} \label{SSM}
\d s^2=A(r)\d t^2-\frac{\d r^2}{1-b(r)/r}-r^2(\d\theta^2+\sin^2\theta \d\varphi^2).
\end{equation}

Two independent components of the Einstein equations $G_{\mu\nu}=8\pi T_{\mu\nu}$ (we accept
$c=G=1$) give
\beqar
\label{Ein1} b' &=& 8\pi r^2\rho,
\\
\label{Ein2} \frac{A'}{A} &=& \dfrac{b+8\pi r^3p_r}{r^2(1-b/r)},
\eeqar
where the prime denotes the derivative with respect to $r$. Third equation
\beq
\label{Ein3} p'_r = \dfrac{2}{r}(p_{t}-p_r)-(\rho+p_r)\frac{A'}{2A},
\eeq
where $p_{t}(r)$ is the transversal pressure, follows from the
energy-momentum conservation law $\nabla_\mu T^{\mu\nu}=0$ providing
an integrability of the Einstein equations.

The metric (\ref{SSM}) describes a static spherically symmetric wormhole if the functions $A(r)$ and $b(r)$ satisfy the flare-out conditions
\beqar
& \label{throatCond1} b(r_\text{min})=r_\text{min},\qquad b'(r_\text{min})<1, &\\&
\label{throatCond2} 0<A(r_\text{min}) <\infty,\qquad A'(r_\text{min})<\infty. &
\eeqar
The radial coordinate $r$ increases from a minimum value $r_\text{min}$ to infinity. Without loss in generality we take $r_\text{min}=1$. At this radius the coordinate spheres achieve a minimum area and therefore the surface $r=1$ is interpreted  as a wormhole throat.  In this approach $r$ simultaneously corresponds to two maps covering the spacetime on either side of the throat.

Traversable wormholes also require  that $A(r)$ should be
positive and $b(r)<r$ elsewhere to avoid horizons and
singularities. These requirements are ruled out for the considered spacetime configuration in favor of the horizon located at the radius $r=r_0>1$.

\begin{figure}[t]
\includegraphics[width=.45\textwidth]{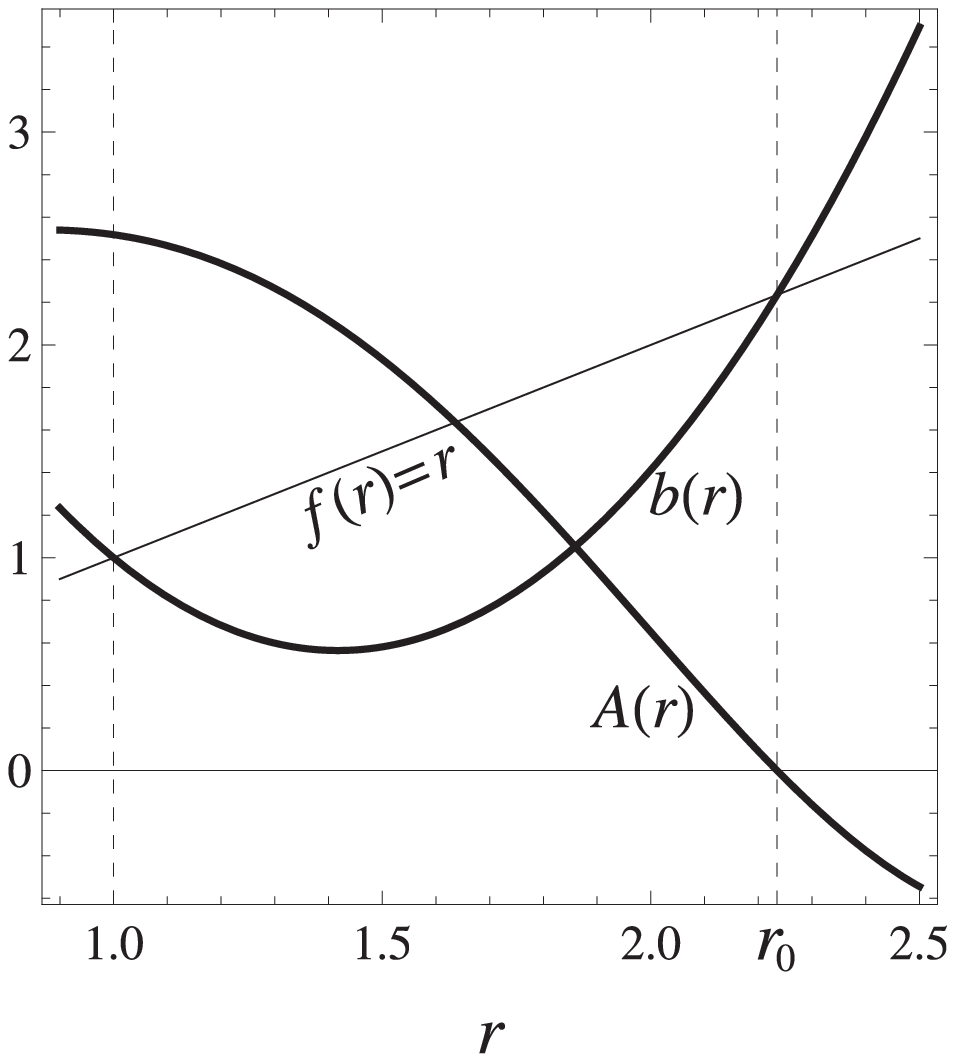}\hfill%
\raisebox{.8\height}{\includegraphics[width=.51\textwidth]{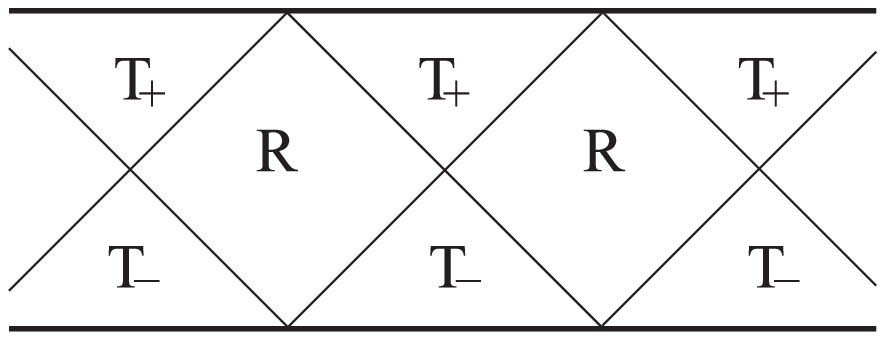}}
\\%
\parbox[t]{0.45\textwidth}{\caption{The functions $A(r)$ and $b(r)$ for the geometry with a wormhole at $r=1$ and simple horizons at $r=r_0$.}\label{Fig-AbNeed}}\hfill
\parbox[t]{0.51\textwidth}{\caption{The global Carter-Penrose structure diagram of the model. The thick lines correspond to $r=\infty$ and the thin lines correspond to the horizons at $r=r_0$. }\label{Fig-CPdiag}}
\end{figure}

The horizon corresponds to simultaneous solutions of the equations
$b(r_0)=r_0$ and $A(r_0)=0$ as it depicts in Fig.~\ref{Fig-AbNeed}.
We consider only simple horizons which imply the following
decomposition for the metric functions near $r_0$
\beq
\label{hrznCond1} b(r)=r_0+b'(r_0)(r-r_0)+o(r-r_0),
\eeq
\beq
\label{simpleZero} A(r)\propto (r-r_0)+o(r-r_0). \eeq

In this case $r_0$ corresponds to the simple horizon separating a static R-region at $r<r_0$ from a non-static  T-region at $r>r_0$. Indeed, the square of the normal to a surface $r$=const in the metric (\ref{SSM}) is $\Delta =  b/r-1$, and $\Delta<0$ when $1<r<r_0$. In this region the spacetime is static (R-region) and $r$ is a spatial coordinate. At $r>r_0$ the normal is a spacelike vector and $\Delta>0$. This is a non-static T-region where $r$ becomes a temporal coordinate. It is verified using the Kretschmann scalar  that the class of metrics under consideration is regular elsewhere \cite{Mokeeva} including the horizons. The global structure of the spacetime is shown in Fig.~\ref{Fig-CPdiag}. It is represented as diagrams for de Sitter geometry joined together through the R-regions.

The regularity conditions on the horizon imply a continuous behavior of the pressures and the energy density. When crossing the horizon $\rho$ and $p_r$ interchange their positions in the stress-energy tensor and for their continuity the condition $p_r+\rho=0$ should be fulfilled on the horizon regardless of a matter kind \cite{Bronnikov3}. For the Chaplygin gas it yields
\beq \label{hrznCond3}
\rho(r_0)=-p_r(r_0)= \sqrt\alpha.
\eeq
Note, that the equation of state (\ref{Chap}) is precisely the same in the R- and T-regions. It means that both regions contain identical matter.

Compatibility of the Einstein equations near the horizon also leads to the following relation for the Chaplygin gas
\beq \label{hrznCond2}
b'(r_0)=8\pi r_0^2 \sqrt\alpha.
\eeq
and provides the continuity of the transversal pressure that takes the value $p_t(r_0)=b''(r_0)/8\pi r_0-3\sqrt\alpha$ on the horizon.

The equation of state (\ref{Chap}) relates the energy density with the radial pressure only. It allows to solve the problem predetermining the metric function $b(r)$ satisfying the flare-out and horizon conditions. The other functions are found from Eqs.~(\ref{Ein1})--(\ref{Ein3}) and produce different cosmological scenarios in the T-regions which are presented in Sec.~\ref{Sec-ExactSolutions}.

\section{A wormhole between two horizons}
\label{Sec-Static region}

When $r$ increases from 1 to $r_0$ the metric (\ref{SSM}) describes a static R-region. To appreciate its structure it is convenient to transform the  metric (\ref{SSM}) to the form
\beq\label{GuassMetr}
\d s^2 = B(x) \d t^2 - \d x^2 - r^2(x)(\d\theta^2+\sin^2\theta \d\varphi^2),
\eeq
where $r(x)$ is an inverse function with respect to the proper length
\beq\label{r in R 2}
x = \int \frac{\d r}{\sqrt{1-b(r)/r}},
\eeq
and $B(x)=A(r(x))$.

To give the simple description of this region, consider the function  $b(r)=r(r-1)(r-r_0)+r$. In this case the coordinate sphere radius has the simple form
\beq\label{Rcos}
r(x) = \frac{1}{2}\left((r_0+1) -(r_0-1)\cos x\right),
\eeq
where $x$ lies in the range $[-\pi,\pi]$ and
\beq\label{RB}
B(x) = B_0\,\frac{r_0-r}{r^{1-1/r_0}}\,
      \frac{
e^{-(r_0+1)^2(2r_0-1)(r_0^2-r_0+1)
                    \text{arctan}\left((3r-r_0-1)/\sqrt{2+r_0-r_0^2}\right)/3r_0^4\sqrt{2+r_0-r_0^2}}
         }{\left(3r_0^2-2(r_0-1)r+r_0+1\right)^{(r_0^3+1)(2r_0^2-2r_0-1)/6r_0^4(r_0-2)}}\, ,
\eeq
where $r$ is taken from (\ref{Rcos}) and $r_0$ is restricted by $1<r_0<2$.

When $x\to\pm\pi$, the function $B(x)$ becomes zero
providing the evidence for horizons which are spaced at a finite
distance apart for a static observer. The function $r(x)$ has a
minimum at $x=0$ implying a wormhole throat, so the
R~region is a spherically symmetric one with the wormhole
surrounded by two horizons.

\begin{figure}[t]
\includegraphics[width=.51\textwidth]{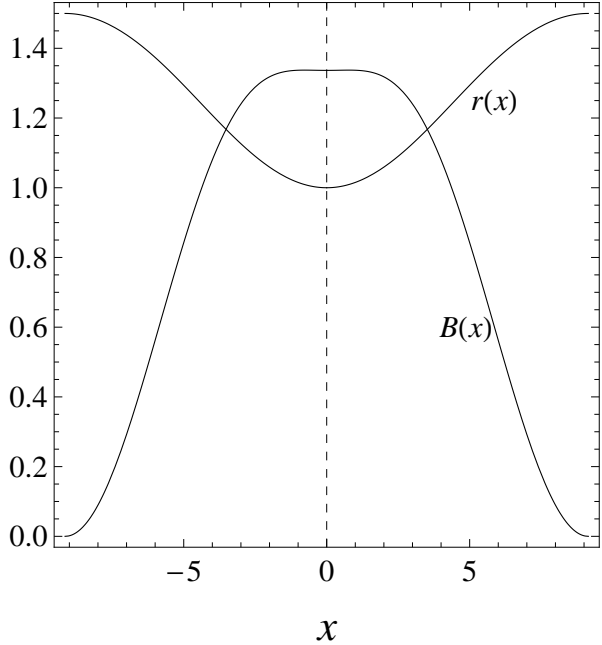}\\%
\parbox[t]{0.51\textwidth}{\caption{The metric functions depending on the proper length $x$ in the R-region.}\label{Fig-Rmetrics2}}
\end{figure}

For an arbitrary function $b(r)$ based on the conditions (\ref{throatCond1}) and (\ref{hrznCond1}) the qualitative behaviors of the metric functions $r(x)$ and $B(x)$ are similar to (\ref{Rcos}) and (\ref{RB}), and the typical behavior of $r(x)$ and $B(x)$ are depicted in Fig.~\ref{Fig-Rmetrics2}.

\section{Kantowski-Sachs cosmologies}
\label{Sec-ExactSolutions}

\subsection{General properties}

The metric (\ref{SSM}) in the T-region can be rewritten as
\beq \label{KSmetr}
\d s^2= \frac{\d\eta^2}{b(\eta)/\eta-1} - \tilde{A}(\eta)\d \rho^2 - \eta^2\d\Omega^2,
\eeq
where $\tilde{A}(\eta)=-A(\eta)$ is the positive function for $\eta>r_0$. We introduce the new coordinates, $\eta$ instead $r$, and $\rho$ instead $t$, to mark the interchange of roles between the spatial and temporal coordinates. The integral
\beq \label{PhysTime}
\tau = \pm\int \frac{\d\eta}{\sqrt{b(\eta)/\eta-1}}
\eeq
defines the physical time
$\tau$ for an observer in the T-region. The Carter-Penrose diagram in Fig.~\ref{Fig-CPdiag} shows both the contracting region T$_-$ corresponding to $\tau<0$ and the expanding region T$_+$ corresponding to $\tau>0$. The metric (\ref{KSmetr}) describes a Kantowski-Sachs (KS) cosmological model \cite{KS}. The spatial part of the KS universe (i.\,e. surfaces $\eta=$const) has the spherical symmetry but it is highly anisotropic. A topology of the spatial section is $\mathbb{R}\times\mathbb{S}^2$ and a three-dimensional cylinder expands with two scale factors for the spherical and longitudinal directions which depend on the proper time $\tau$ according to $a_t(\tau)=\eta(\tau)$ and $a_r(\tau)=\tilde{A}^{1/2}\bigl(\eta(\tau)\bigr)$  correspondingly.

In all cases the evolution starts with a NBB from a horizon at $\eta=r_0$. There is a coordinate singularity which the observer interprets as the beginning of the universe. It starts to expand anisotropically with the initial values of the scale factors $a_r=0$ and $a_t=r_0$. The later evolutions are quite different and several exact solutions corresponding to different scenarios were obtained in \cite{Mokeeva}. The present objective is to find a spacetime configuration providing isotropization at late times to be of interest in cosmological context.

Since the spherical  scale factor $a_t$ linearly increases with the coordinate time $\eta$,  asymptotical equalizing the scale factors implies that the metric function $b(\eta)$ becomes monotonic one from some instant so that $b(\eta)\propto\eta^\beta$ for $\eta\to\infty$. It provides a monotonic behavior of the longitudinal scale factor. Before this stage the evolution, in principle, may be arbitrary including, for example, cyclic or quasi-cyclic regimes. It is readily seen in Fig.~\ref{Fig-AbNeed} that the rate of growth  $\beta\ge 1$ to avoid singularities in the T-region.
All the scenarios can be divided into three types according to a value of $\beta$. The scenario of the first kind corresponds to $\beta<3$, the second kind does $\beta>3$ and the third one is realized for $\beta=3$. All these types are considered below.

\subsection{Type 1 scenarios: eternal anisotropic expansion}

First we consider the case of $\beta<3$. It follows from the equations (\ref{Chap}) and (\ref{Ein1}) that the dominant term in the large $\eta$ expansion of rhs in the equation (\ref{Ein2})  is $\eta^{5-2\beta}$. It leads to principally exponential growth of the function $A(\eta)$ with respect to the coordinate time. The integral (\ref{PhysTime}) for large $\eta$ gives the power dependence between the physical and coordinate times, $\tau\propto\eta^{(3-\beta)/2}$. Due to this asymptotic behavior the universe evolution, having started at some instant in the past for the KS observer, lasts for  infinite time. All this time the universe anisotropically expands. In the remote future the scale factor along the coordinate spheres increases with a power of the proper time $\tau$  while the scale factor in the longitudinal direction grows in the mixed manner, power and exponential:
\beq\label{asym<3}
a_r\propto\tau^{\kappa}\exp\tau^4, \qquad a_t\propto\tau^{2/(3-\beta)},
\eeq
where the constant $\kappa$ is determined by an explicit form of $b(\eta)$.

Asymptotic behavior of the pressures and the energy density can be estimated from the equations (\ref{Chap}), (\ref{Ein1}) and (\ref{Ein3}). In doing so it should be kept in mind that the energy density  and  the radial pressure interchange their roles when crossing the horizon. In accordance with an asymptotic behavior of $b(\eta)$ we find different dependencies on the coordinate time which transform to the $\beta$-independent behavior for the proper time. From observer's point of view the radial pressure decreases in its absolute  value as $p_r\propto\tau^{-2}$ and the main contribution in the energy balance is given by the energy density growing quadratically in the proper time, $\rho\propto\tau^2$,  and the negative transversal pressure which increases in its absolute value as $p_t\propto\tau^6$.

To demonstrate the exact solution let us take the metric function $b(\eta)$ in the form (the growth rate $\beta=1$)
\beq\label{b 1}
b(\eta)=\frac{d}{\eta}(\eta-1)(\eta-r_0)+\eta.
\eeq

The latter metric function is found as
\beq\label{A 1}
A(\eta)=A_0 \frac{\eta-r_0}{\eta} \left(\eta-q_0\right)^{q_5+q_6} \left(\eta+q_0\right)^{q_5-q_6} \exp\left\{\frac{q_1}{4}\eta^4+\frac{q_2}{3}\eta^3+\frac{q_3}{2}\eta^2+q_4\eta\right\},
\eeq
where the constants
\beqar
&\displaystyle
d=\frac{r_0^3\sqrt{r_0^6+4r_0+4}-r_0^6-2r_0}{2(r_0-1)},
&\\&\displaystyle
q_0=\sqrt{\frac{r_0d}{d+1}},\qquad
q_1=\frac{(d+1)(r_0^2+q_0^2)}{r_0^4d},\qquad
q_2=q_1(r_0+1),
&\\&\displaystyle
q_3=q_1\left(r_0^2+2q_0^2+q_0^2/d+1\right),\qquad
q_4=q_1(r_0+1)(r_0^2+q_0^2+1),
&\\&\displaystyle
q_5=\frac{q_0^6(2d+1)}{r_0^5d(1-q_0^2)},\qquad
q_6=\frac{q_0^5(r_0+1)}{r_0^5(1-q_0^2)}
\eeqar
are positive and determined by the horizon radius $r_0$ only. Besides, the solution implies $q_0<1$, so the function $A(\eta)$ is zero only at the horizon and positive for $\eta>r_0$.

\begin{figure}[t]
\includegraphics[width=0.42\textwidth]{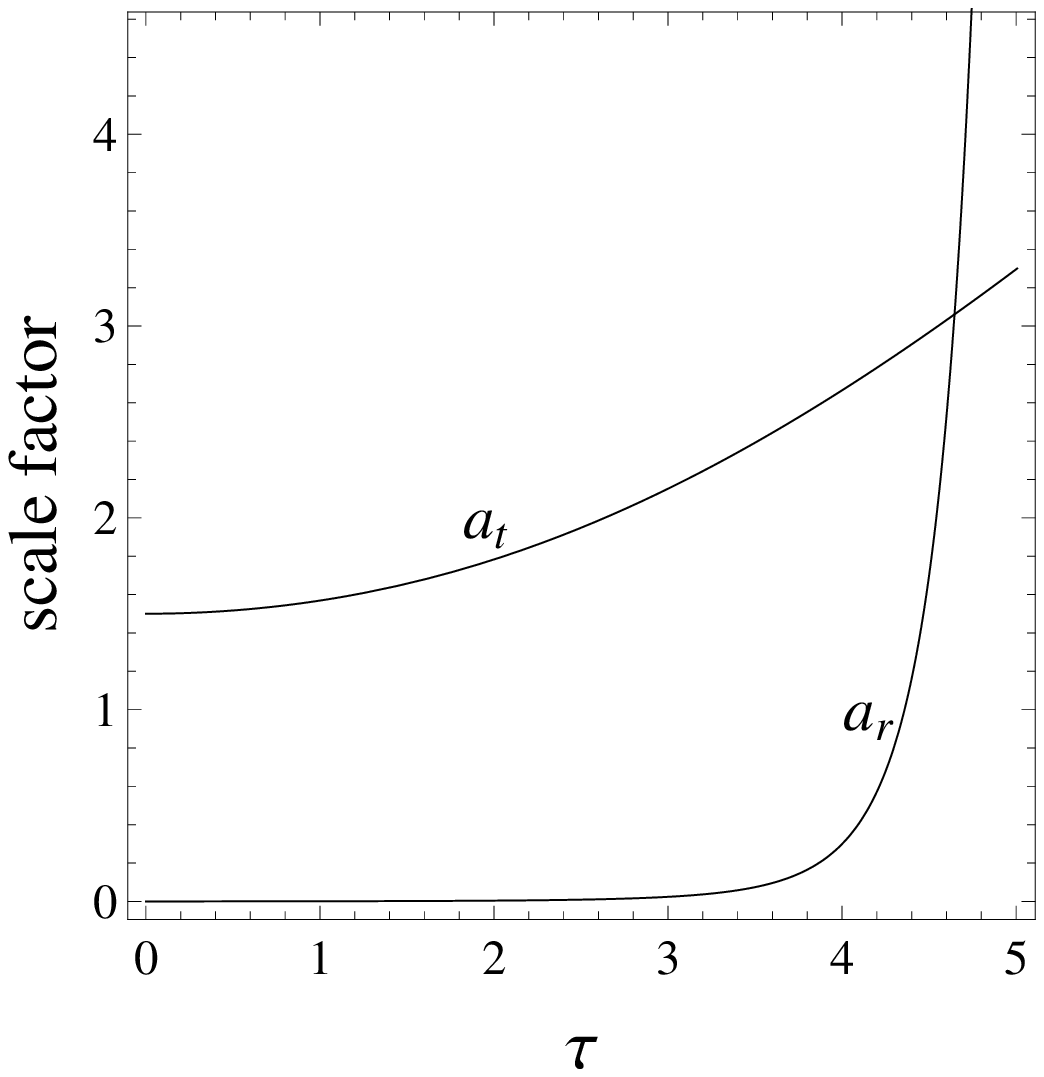}\hfill%
\includegraphics[width=0.52\textwidth]{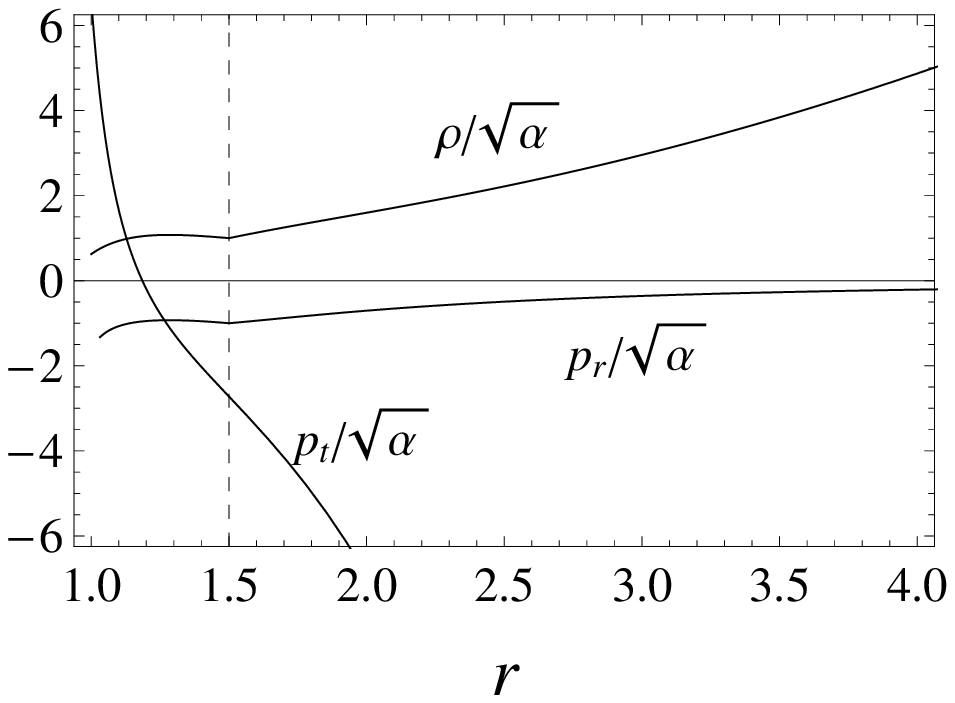}\\
\parbox[t]{0.42\textwidth}{\caption{The increase of the scale factors for the solution specified by Eqs.~(\ref{b 1}) and (\ref{A 1}) for  $r_0=1.5$. }\label{Fig-Tmetrics1}}\hfill
\parbox[t]{0.52\textwidth}{\caption{\label{Fig-prho1}
The normalized energy density, radial and transversal pressures as
functions of the radial coordinate for the type 1 scenario. The
dotted line separates the R- and T-regions.}}
\end{figure}

In this solution the physical time $\tau$ varies directly as the coordinate one when $\eta\to\infty$. At late times the asymptotic expansion occurs according to (\ref{asym<3}) with $\kappa=q_5$ and $\beta=1$. Figs.~\ref{Fig-Tmetrics1} and \ref{Fig-prho1} give a pictorial representation of the scenario considered.

\subsection{Type 2 scenarios: universes with finite lifetime}

The second type of scenarios includes  the functions $b(\eta)$ with the asymptotical growth rate $\beta>3$. In this case for large $\eta$ the equation (\ref{Ein2}) yields $A'/A\propto-1/\eta$ from whence it follows that $A(\eta)\propto\eta^{-1}$. It is also easy to see from Eq.~(\ref{PhysTime}) that if the function $b(\eta)\to\infty$ faster than $\eta^3$  for $\eta\to\infty$, then the physical time has a final point. These reasons lead to the following qualitative pattern of the universe evolution.

%
\begin{figure}[t]
\includegraphics[width=0.42\textwidth]{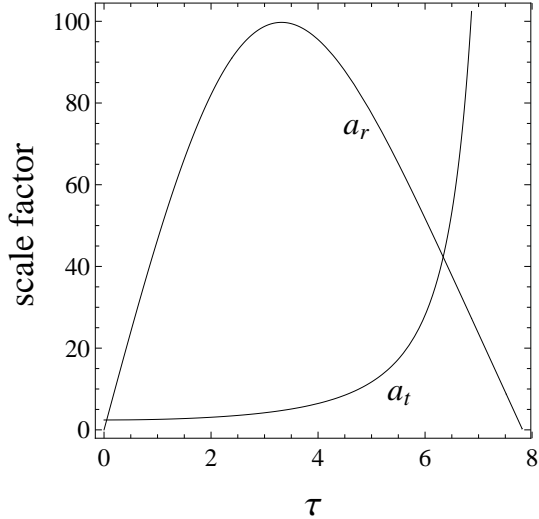}\hfill%
\\
\parbox[t]{0.42\textwidth}{\caption{The scale factors for the universes with finite lifetime.}\label{Fig-Tmetrics4}}\hfill
\end{figure}

The expansion in the T-region starts with a NBB and the ensuing cosmological evolution develops during finite cosmological lifetime from the point of view of the KS observer. Throughout this period the spherical scale factor infinitely increases and the longitudinal scale factor peaks at some instant and falls to zero as it is shown in Fig.~\ref{Fig-Tmetrics4}. It easy to see from the asymptotic dependence of the scale factors that at late stage $a_r\propto a_t^{-1/2}$. Note, that the contraction in the longitudinal direction occurs even though both, radial and transversal, pressures are negative and the energy density disappears  towards the end of the evolution. This is typical for the KS model where a spherical acceleration is determined by the radial pressure and the spherical Habble parameter $H_t=\dot a_t/a_t$, where the dot denotes the derivative with respect to the proper time $\tau$, while a longitudinal acceleration also depends on the transversal pressure and the energy density according to
\beqar
&\displaystyle
\frac{1}{a_t^2}+H_t^2+2\frac{\ddot a_t}{a_t}=-8\pi p_r,
&\\&\displaystyle\label{radialAcceleration}
\frac{1}{a_t^2}+H_t^2-\frac{\ddot a_r}{a_r}=4\pi \left( \rho -p_r +2p_t\right).
&
\eeqar
If the energy density is neglected and the rate of growth of $H_t$ is slight so that $|p_r|>2|p_t|+H_t^2/4\pi$, then one observes a decelerated expansion or contraction  in the longitudinal direction.

The metric functions
\beqar
b(\eta) &=& 4(3-2\sqrt 2)\eta^4-11+8\sqrt 2,\label{b 4}\\
A(\eta) &=& A_0\frac{
(\eta-1-\sqrt{2}) e^{
-(1+\sqrt{2})(2+\sqrt{2}) \text{arctan} \left( \left(2\eta+2+\sqrt{2}\right)/\sqrt{14+8\sqrt{2}} \right) /\sqrt{14+8\sqrt{2}}
}   
}{\eta\sqrt{\eta^2+(2+\sqrt{2})\eta+2+3\sqrt{2}}}\label{A 4}
\eeqar
give the example of an exact solution associated with the scenarios
of the concerned type ($\beta=4$) with the horizon radius
$r_0=1+\sqrt 2$. Fig.~\ref{Fig-Tmetrics4} depicts the scale factors
produced by the metric functions (\ref{b 4}) and (\ref{A 4}).

\subsection{Type 3 scenarios:  a quasi de Sitter late stage}

Consider now the case of $\beta=3$. At the late stage the relation
between the coordinate and physical times is a pure expononetial and
the equation (\ref{Ein2}) implies the solution
$A(\eta)\propto\eta^\gamma$, where the constant $\gamma$ depends
from an explicit form of $b(\eta)$. The scenario for the KS observer
is following. The universe starts its evolution from the horizon and
lasts during for infinite time. The asymptotical behavior of the
scale factors for the longitudinal and lateral directions as
functions of the proper time when $\tau\to\infty$ is
\beq\label{asym=3}
 a_r\propto\exp h_r\tau, \qquad a_t\propto\exp
h_t\tau,
\eeq
where the constants $h_r$ and $h_t$  are the
asymptotic Habble parameters in the longitudinal and spherical
directions respectively. The constant $h_t$ is positive and the
constant $h_r$ can be positive, negative and zero. In accordance
with the sign, the initial longitudinal expansion can continue, give
place to the contraction or settle on a constant value at the remote
future. In any case the energy density and the pressures tend to
constants. This regime is very similar to the de Sitter state and,
moreover, the anisotropic evolution can isotropize if the metric
functions admit the solution with $h_r=h_t$.

To study this type of evolution in more details, let us use the
metric function
\beq\label{b 3}
b(\eta)=d(\eta-1)(\eta-r_0)(\eta-l)+\eta,
\eeq
where the constants
$d>0$ and $l<1$. Solving the equations (\ref{Chap})--(\ref{Ein3})
under the conditions discussed in the Sec.~\ref{Sec-Conditions} one
obtains the metric function $A(\eta)$ in one of two forms
\beq\label{greenA}
A(\eta)=A_0\frac{(\eta-r_0)(\eta-l)^{q_1}%
\exp\left\{q_3\, \text{arctan}\left(q_4\left[\eta-(1+r_0+l)/3\right] \right)\right\}
}{\eta%
\left(\eta^2 -2(1+r_0+l)\eta/3d +r_0+l+r_0l+1/d \right)^{q_2},%
} \eeq or \beq\label{redA}
A(\eta)=A_0\frac{(\eta-r_0)}{\eta}(\eta-l)^{q_1}(\eta-q_5)^{q_7}(\eta-q_6)^{q_8}
\eeq depending on the way that the rational expression in the rhs of
(\ref{Ein2}) is represented as a sum of terms with minimal
denominators. Under the conditions (\ref{throatCond2}) the constant
\beq
l=\frac{2+r_0^4-2r_0d-r_0^2\left(-2d+\sqrt{r_0^4+8+4d(r_0-1)^2}\right)}{2d(r_0-1)}
\eeq
is expressed by two parameters, $r_0$ and $d$, as well as all
the constants $q_i$, which are not represented here in details
because of their unhandiness.

\begin{figure}[t]
\includegraphics[width=0.5\textwidth]{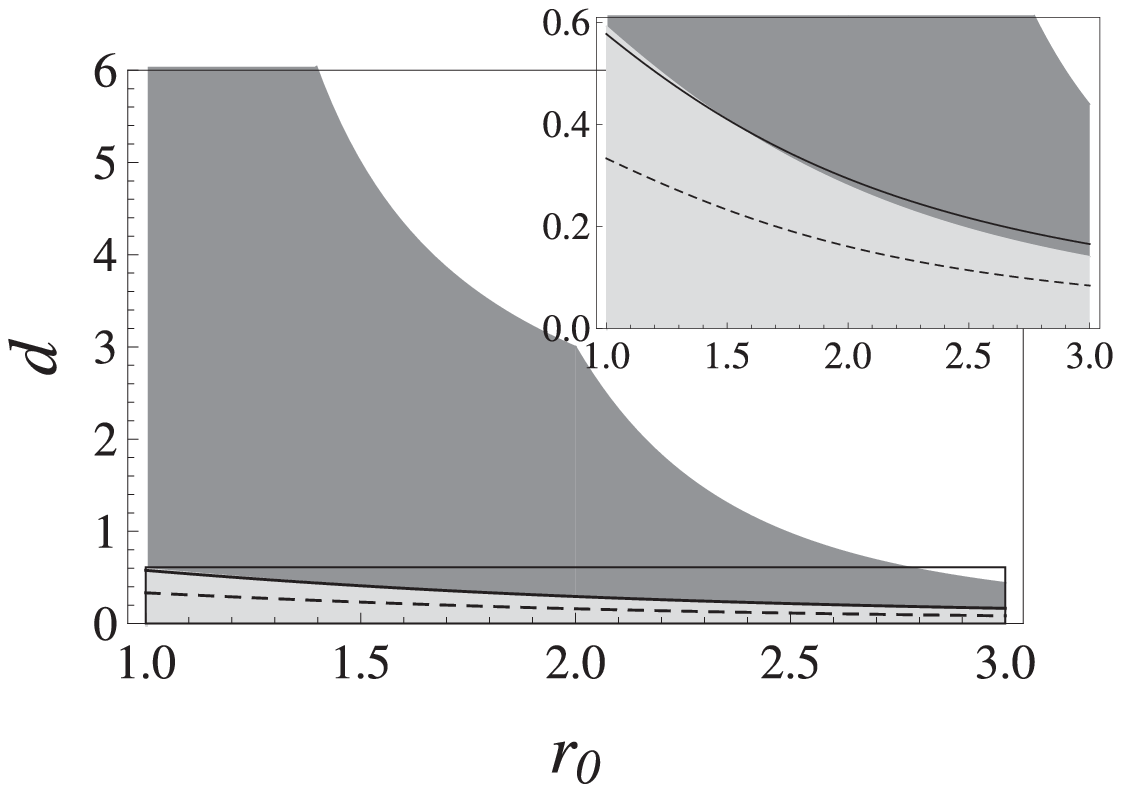}\hfill%
\includegraphics[width=0.45\textwidth]{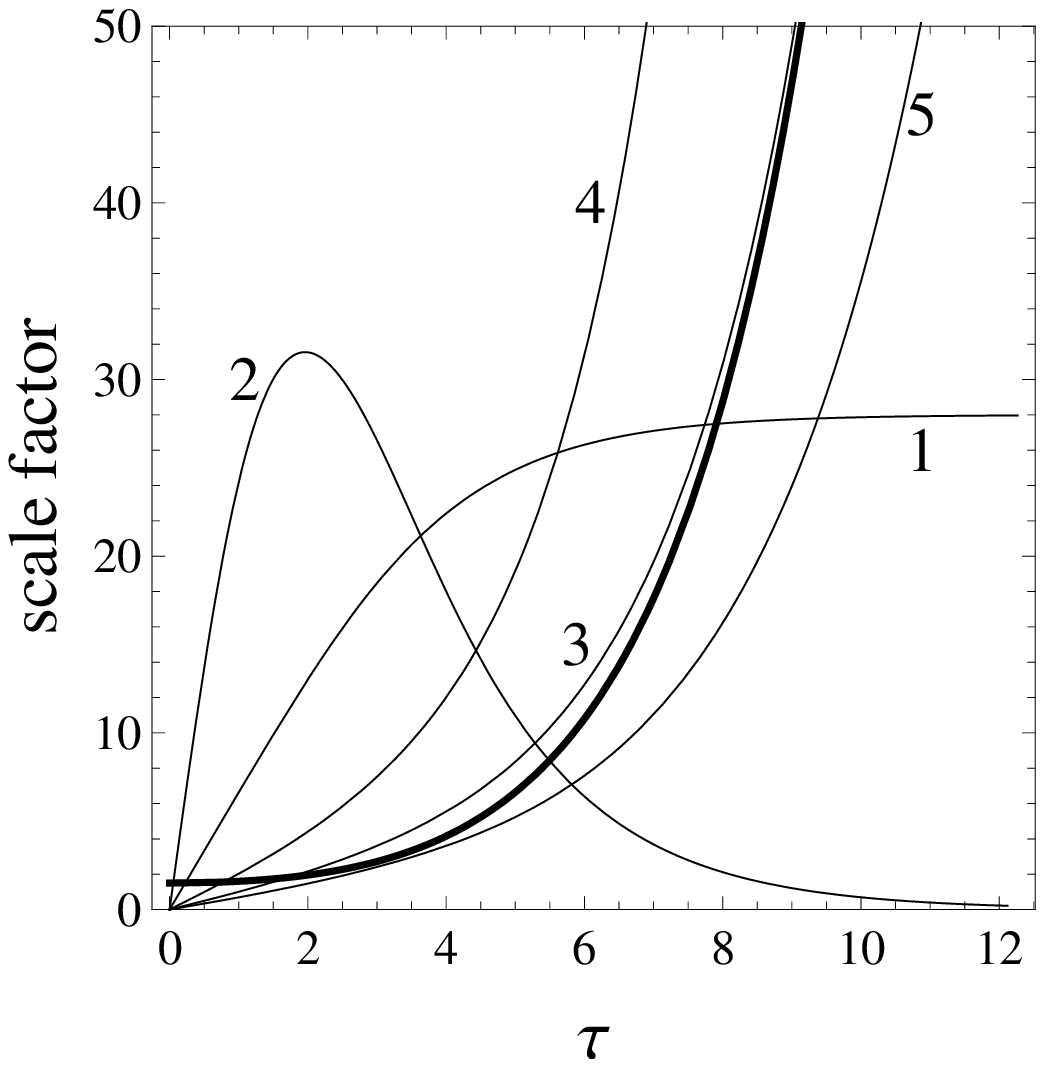}\\
\parbox[t]{0.5\textwidth}{\caption{The parametric diagram represents different sorts of the solutions with a quasi de Sitter behavior at late stage. The area restricted by the rectangle is represented in the top right panel on an enlarged scale. The detail description is in the text.}\label{Fig-ad}}\hfill
\parbox[t]{0.45\textwidth}{\caption{The scale factors in type 3 scenarios for $r_0=1.5$. The thick line corresponds to the spherical scale factor $a_t$. The thin lines represent the longitudinal scale factor $a_r$.
\newline
Line 1: $\gamma=0$, $d\approx0.41$, freezing evolution.
Line 2: $d=2$, $\gamma<0$, contracting universe.
Line 3: $\gamma=2$, $d\approx0.233$, isotropization.
Line 4: $d=0.21$, $\gamma>2$, anisotropic expansion.
Line 5: $d=0.25$, $0<\gamma<2$, idem.
}\label{Fig-Tmetrics3} }
\end{figure}

Singularities and additional horizons do not appear in the solutions if $l<1$. This condition restricts the admitted region for the parameters $r_0$ and $d$  that is shaded in Fig.~\ref{Fig-ad}. The dark grey corresponds to the solution (\ref{greenA}) and the light gray does (\ref{redA}).

As it is seen from (\ref{PhysTime}) at late stage $\eta\propto\exp\sqrt{d}\tau$ and hence, $h_r=\gamma\sqrt{d}/2$, $h_t=\sqrt{d}$ with
\beq
\gamma=\frac{4+r_0^4-4r_0d+2d-6d^2-r_0^2\left(-2d+\sqrt{r_0^4+8+4d(r_0-1)^2}\right)}{6d^2}.
\eeq
Various values of $r_0$ and $d$ bring about different kinds of the quasi de Sitter evolution. These are shown in Fig.~\ref{Fig-ad} and the corresponding behavior of the scale factors is depicted in Fig.~\ref{Fig-Tmetrics3}.

The value $\gamma=0$ represents the solutions describing freezing of
the longitudinal expansion with the scale factor $a_r$ tending to a
constant. The solid line in Fig.~\ref{Fig-ad} corresponds to these
solutions. The parameters over the line produce scenarios with
universes contracting in the longitudinal direction after the
initial expansion. In this region$\gamma<0$. The region under the
line corresponds to solutions with an infinite expansion. The dotted
line in this region is specified by $\gamma=2$ and represents the
parameters providing isotropization at late times. The parameters
under the line generate solutions where the longitudinal expansion
is faster then the spherical one, and the parameters between the
solid and dotted lines give  inverse expansion rates. Solutions of
all these kinds are illustrated in Fig.~\ref{Fig-Tmetrics3}.

An isotropic regime is realized at the late stage if $\gamma=2$ or,
what is the same,
\beq\label{HubbleIsotr}
d=\frac{1}{18}\left(
\sqrt{4r_0^4+8r_0^3-4r_0+73}-2r_0^2-2r_0+1\right).
\eeq
In this case
the universe falls into the  de Sitter mode where directional Hubble
parameters, $h_r$ and $h_t$, reduce to the same value
$H_\infty=\sqrt d$ and both pressures become equal $p_r = p_t =
-3H_\infty^2/8\pi$. This solution is shown in
Fig.~\ref{Fig-Tmetrics3} by the line 3. It is governed by one
independent parameter $r_0$ which dictates the rate of inflation at
the last stage.

Driving parameters can increase in number as well as the scenario can become more complicated when used different functions $b(\eta)$ with the asymptotic behavior $b(\eta)\sim H_\infty^2\eta^3$. The condition specifying the isotropic regime follows from the equations (\ref{Ein2}) and (\ref{hrznCond2}):
\beq
b'(r_0)=3H_\infty^2 r_0^2.
\eeq
As before, it directly relates the de Sitter phase and the NBB.

\section{Conclusion}
\label{Sec-Conclusion}

\begin{figure}[t]
\includegraphics[width=0.42\textwidth]{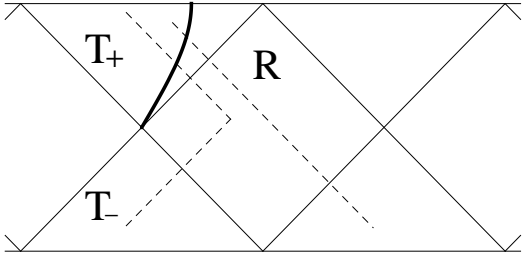}\hfill\\%
\parbox[t]{0.42\textwidth}{\caption{The Carter-Penrose diagram of the model considered with world lines of a motionless KS observer (thick line) and photons (dashed lines) coming from a contracting universe through a wormhole.}\label{Fig-Signals}}\hfill
\end{figure}

In conclusion we summarize the results obtained and discuss some questions concerned.

We have studied  a spacetime with an anisotropic fluid obeying the Chaplygin equation of state. The spacetime consists two T-regions separated by a static spherically symmetric R-region with a wormhole providing a traverse between two horizons. In the T-region an anisotropic KS cosmology is realized. The peculiar feature of  all scenarios is that the evolution starts from the horizon appearing as big bang for a KS observer. This singularity is apparent since the total spacetime is regular. We have focused our attention on the universes with monotonic asymptotic behavior to extract a class of metrics admitting an isotropic late stage.  We have found three types of such cosmologies. The first one describes an universe with an ineradicable anisotropic expansion. The second one represents the contracting universe with finite lifetime. The third one in turn includes three sorts of evolution following the longitudinal scale factor that can increase, decrease or become constant. The connecting feature for these cosmologies is the exponential behavior of the scale factors at late times. The isotropic regime occurs in this case only. When $b(\eta)\sim H_\infty^2\eta^3$  the universe is going into the inflation phase with the Hubble parameter $H_\infty$.

It was shown in \cite{Bronnikov2} that in any universe with a NBB a motionless KS observer can get access to prehistoric information from a static region coming with photons and particles which have crossed the horizon. It means, as applied to our model, that being in an expanding universe T$_+$ one can observe not only a static region but receive signals from a contracting universe T$_-$. We illustrate it using the Carter-Penrose diagram in Fig.~\ref{Fig-Signals}. The thick line gives a motionless observer trajectory that coincides with a geodesic $\eta=\eta(\tau)$ determining by the integral (\ref{PhysTime}). The proper time $\tau$ is affine parameter along the geodesic. The dashed lines correspond to photon trajectories. The straight line corresponds to the photon having crossed two spatially separated horizons and polygonal path does the photon having crossed twice the same horizon after reflection in the R-region.

In closing we estimate a size of the static region for the model with inflation. The asymptotic Habble parameter, $H_\infty$, is related with the horizon radius $r_0$ through the equation (\ref{HubbleIsotr}). Turning back to the dimensional quantities we obtain $H_\infty=c/\sqrt 3 r_0$ when $r_0$ is close to the wormhole throat radius $r_\text{min}$, and $H_\infty=c/ r_0$  when $r_0\gg r_\text{min}$. According to present conceptions the Habble parameter for initial inflation takes its value between $10^{36}$ and $10^{42}$ sec$^{-1}$. It is easy to find that the size of the static region (and the wormhole throat also) falls into the broad range from ten to ten billions of the Plank length. This estimation is in agreement with the classical approach undertaken in the work.

;

\end{document}